\documentclass{aastex631}
\usepackage{float}

\usepackage{amsmath}
\usepackage{subfigure}
\usepackage{multirow}

\begin{document}

\title{An Ultra-Fast Image Simulation Technique with Spatially Variable Point Spread Functions}

\correspondingauthor{Peng Jia}
\email{robinmartin20@gmail.com}

\author{Zeyu Bai}
\affiliation{College of Electronic Information and Optical Engineering, Taiyuan University of Technology, Taiyuan, 030024, China}

\author[0000-0001-6623-0931]{Peng Jia}
\affiliation{College of Electronic Information and Optical Engineering, Taiyuan University of Technology, Taiyuan, 030024, China}

\author{Jiameng Lv}
\affiliation{College of Electronic Information and Optical Engineering, Taiyuan University of Technology, Taiyuan, 030024, China}

\author{Xiang Zhang}
\affiliation{College of Electronic Information and Optical Engineering, Taiyuan University of Technology, Taiyuan, 030024, China}

\author{Wennan Xiang}
\affiliation{College of Electronic Information and Optical Engineering, Taiyuan University of Technology, Taiyuan, 030024, China}

\author{Lin Nie}
\affiliation{Department of Information Engineering, Wuhan Institute of City, Wuhan, Hubei 430083, China}



\begin{abstract}
Simulated images are essential in algorithm development and instrument testing for optical telescopes. During real observations, images obtained by optical telescopes are affected by spatially variable point spread functions (PSFs), a crucial effect requiring accurate simulation. Traditional methods segment images into patches, convolve patches with individual PSFs, and reassemble them as a whole image. Although widely used, these approaches suffer from slow convolution processes and reduced image fidelity due to abrupt PSF transitions between different patches. This paper introduces a novel method for generating simulated images with spatial continuously varying PSFs. Our approach firstly decomposes original images into PSF bases derived with the principal component analysis method. The entire image is then convolved with these PSF bases to create image bases. Finally, we multiply the coefficients of image bases with these image bases for each pixels and add the multiplication results along each pixel to obtain the final simulated image. Our method could generate high-fidelity simulated images with spatially variable PSFs without boundary artifacts. The method proposed in this paper significantly improves the speed of astronomical image simulation, potentially advancing observational astronomy and instrumental development.
\end{abstract}



\section{Introduction} \label{sec:intro}
Simulated observational data is a crucial tool in modern astronomy, serving as a cornerstone for the development and evaluation of innovative astronomical instruments and data processing techniques. The ability to precisely control the quality and content of simulated data provides researchers with an invaluable means to assess data processing methods and design new astronomical instruments \citep{le2006adaptive, carbillet2008software, connolly2010simulating, liu2011simulation, wang2012computer, le2012simulations, van2014wide, perrin2014james, briguglio2022simulator}. Simulated images, in particular, prove exceptionally useful for testing image processing algorithms and forecasting scientific outcomes for a particular scientific project \citep{jee2011toward, penny2013exels, hoekstra2017study, lanusse2018cmu, korytov2019cosmodc2, huber2019strongly, sanchez2020lsst, schmitz2020euclid}. In recent years, deep learning based data processing approaches have been widely discussed. Since most deep learning algorithms need training data with well distribution as prior information, simulated images can be used to significantly improve the efficiency and accuracy of the deep learning based approach \citep{buchanan2022gaussian, yue2022mock, scaramella2022euclid, zhou2022extracting, cao2022anisotropies, liu2022potential, jia2022detection, he2022galaxy, song2022synergy, deng2022forecasting, li2022searching, shen2022tolerance}. Furthermore, simulated images play a pivotal role in the development of adaptive optic systems and different image restoration algorithms. These applications require a substantial number of high-fidelity simulated images for a comprehensive performance evaluation \citep{mackay2013high, denker2018image, lu2018ground, rodeghiero2018impact, mieda2018multiconjugate, jia2019solar, beltramo2019prime, fetick2020blind, monty2021towards, wang2021blind, jia2021data, lauritsen2021superresolving, femenia2022adaptive, minowa2022ultimate, nammour2022shapenet, guyon2022high, hitchcock2022thresher, bernardi2022restoration, yang2023correction}. \\

The diverse applications of simulated images underscore their fundamental importance in driving progress across multiple domains of astronomical research and instrumentation. As we continue to push the boundaries of our observational capabilities, the role of high-quality simulated data in refining our techniques and instruments becomes increasingly critical. Researchers typically follow a systematic procedure to generate simulated images, which can be outlined as follows:
\begin{itemize}
\item Obtain Point Spread Function (PSFs): Acquire PSFs for various fields of view or different observation times, based on the optical design and aberration properties of the imaging system.
\item Convolve Original Images: Convolve segments of high-resolution original images with corresponding PSFs to create 'convolved images' that account for optical system blurring effects.
\item Generate Optical Images: Use the 'convolved images' as prior probability distributions for photons, employing random number generators to simulate photon redistribution and create 'optical images'.
\item Add Noise and Sampling Errors: Incorporate various noises and sampling errors into the 'optical images' to simulate detector properties and inherent noise in astronomical imaging systems.
\end{itemize} 

In the processes described above, the PSF plays a crucial role in generating simulated astronomical images. The PSF characterizes how a point source of light is dispersed by an optical system, encompassing various real effects. These include static aberrations arising from optical design or manufacturing imperfections, as well as dynamic aberrations caused by atmospheric turbulence or misalignment of optical components \citep{piotrowski2013psf, la2015method, wagner2019overview, beltramo2020review}. However, the computation of PSFs and their subsequent convolution with "original images" to produce "optical images" poses several significant challenges. 
\begin{itemize}
\item PSFs typically exhibit variations across different fields of view in real astronomical observations. However, current simulation techniques often simplify this complexity by assuming that PSFs remain uniform within small image patches. While this approach facilitates easier computation, it represents a departure from the physical reality of optical systems.
\item When simulating large, extended astronomical objects such as galaxies or nebulae, researchers often need to divide the image into smaller patches. Each of these patches is then convolved with PSFs corresponding to different fields of view. However, this piecewise approach can introduce artifacts in the final simulated image, such as visible seams or abrupt transitions between sections. 
\end{itemize} 

To solve the above problems, this paper introduces a novel approach for efficiently generating images with spatially varying PSFs. We employ Principal Component Analysis (PCA) to decompose PSFs into a set of basis functions, allowing PSF in each pixel to be represented by a series of parameters. Our method involves convolving the original image with these PSF bases to produce image bases, which are then weighted by pixel-specific parameters and combined to form the final simulated image. By convolving the entire image with a few PSF bases, rather than numerous individual PSFs in different fields of view, our approach significantly reduces computational time and memory usage. Furthermore, the resulting images exhibit pixel-level PSF variations without boundaries within observation images, providing a more accurate representation of real astronomical observations. The details of our method are presented in Section~\ref{sec:method}, followed by results in Section~\ref{sec:results}, and concluding remarks and future research directions in Section~\ref{sec:con}.\\

\section{Method} \label{sec:method}
\subsection{The Basic Principle} \label{basic}
PSFs in observation images are influenced by two main factors: optical system diffraction and wavefront errors caused by atmospheric turbulence or optical system aberrations. Optical system diffraction leads to the appearance of diffraction rings or similar patterns, while wavefront errors cause PSF deformation. As wavefront errors change continuously, PSFs also vary continuously across different fields of view. Traditional methods simulate images affected by these PSFs through dividing the original image into small patches, each convolved with different PSFs. However, this approach is computationally intensive. For example, a typical $2048 \times 2048$ pixel image divided into $16 \times 16$ pixel patches requires 16384 convolution operations, as each patch corresponds to an operation. Furthermore, these methods can introduce visible boundaries between patches, which may reduce the effectiveness of the simulated images for performance testing in various methods. \\

\begin{figure}[h]
\centering 
\includegraphics[width=10cm]{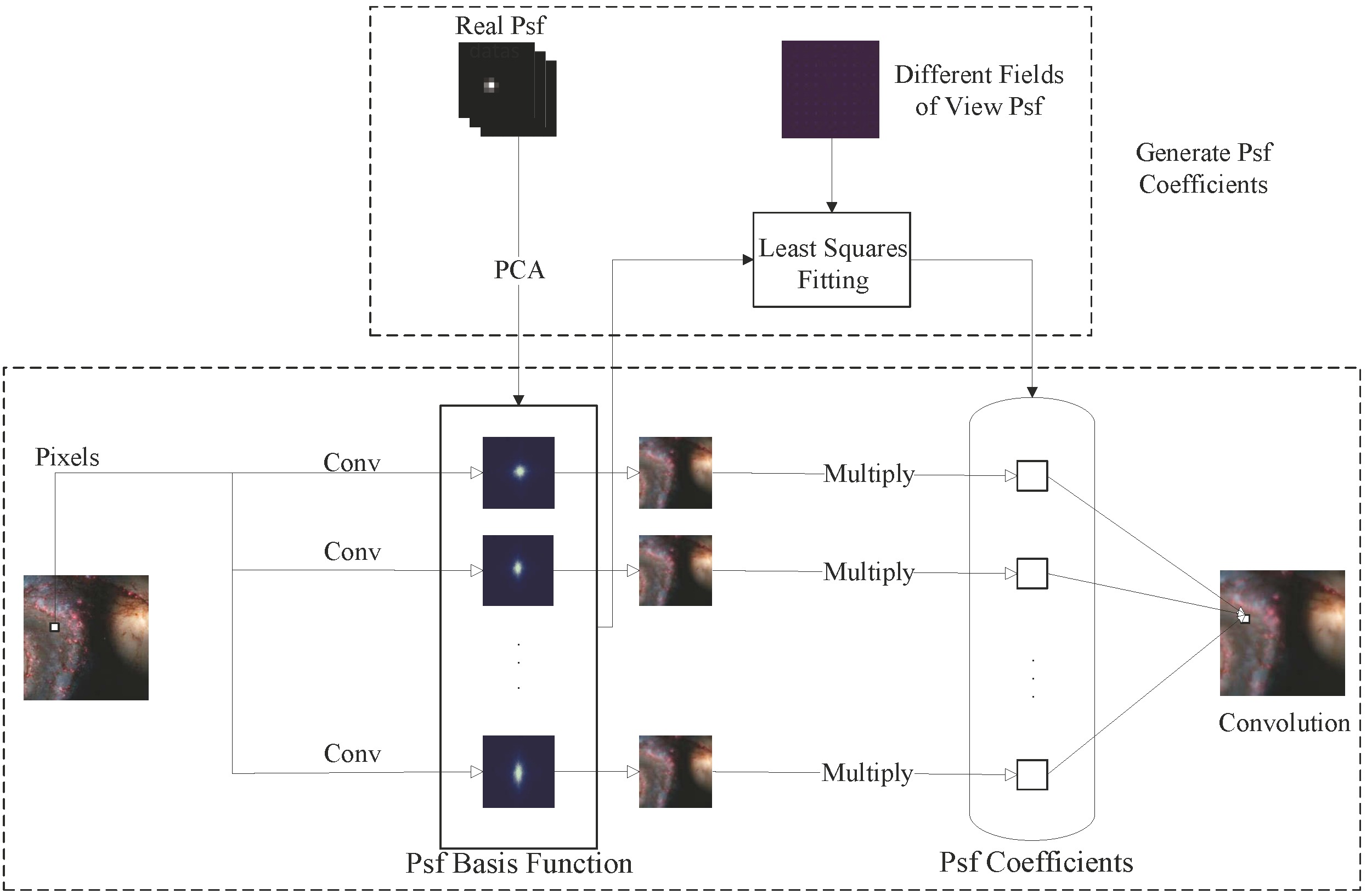} 
\caption{The method proposed in this study is delineated in this figure, which illustrates a two-stage procedure for image simulation. The first stage involves coefficient matrix generation, wherein PSF basis are derived from empirical PSF data. The second stage encompasses the convolution and addition step, where the previously obtained PSF basis functions are convolved with the original image to generate image bases. These image bases are weighted by the coefficients in each pixel and added together to produce the final blurred image. }\label{figure1} 
\end{figure}

It is well-established that PSFs of optical telescopes can be effectively represented using various bases, either through statistical methods or physically-inspired modeling approaches \citep{jee2007principal, krist201120, sun2022precise, liaudat2023point}. These representations allow for a more efficient parameterization of PSF components, which can be applied to cosmic shear measurements or precise astrometry and photometry. It is worth noting that the aforementioned methods are primarily used for data post-processing. While in this paper, we adapt these concepts to generate simulated images affected by spatially variable PSFs by forward modeling as shown in figure~\ref{figure1}. The underlying principle is that by decomposing PSFs into specific bases and performing convolution of images with spatially variable PSFs, we can express this process mathematically as follows:
\begin{equation}
\label{eq1}
FImg = \begin{bmatrix}
FImg_1 \\
\vdots \\
FImg_{N\times N}
\end{bmatrix}
=PSF \ast IImg
=\begin{bmatrix}
PSF_{1}^{T} \cdot IImg \\
\vdots \\
PSF_{N\times N}^{T} \cdot IImg,
\end{bmatrix}
\end{equation}
where $FImg$ represents the final simulated image and $IImg$ denotes the original image. For simplicity, we assume that both the PSF and the original image have the same dimensions of $N\times N$ pixels. In cases where larger image sizes are considered, we can easily adapt the above equations to represent the multiplication of two matrices with different dimensions. If we consider that PSFs can be decomposed into specific bases, we can express this as:
\begin{equation}
\label{eq2}
PSF_{N\times N} = \sum_{m=1}^{M} \alpha_{m,n} \cdot PSFBases_m,
\end{equation}
where $\alpha$ represents the coefficients of $PSFBases$, which are used to represent PSFs using their respective bases. With these representations in place, we can rewrite Equation~\ref{eq1} as follows:
\begin{equation}
\label{eq3}
FImg_{N\times N} = \sum_{m=1}^{M} \alpha_{m,n} \cdot PSFBases_{m}^{T} \cdot IImg,
\end{equation}
where $m$ represents the coefficients of PSF bases convolved with the original image, and $n$ denotes the PSF size. When considering the convolution of the original image with spatially variable PSFs, we would typically need to perform $N\times N$ convolutions (where $N$ is the size of the original image). However, our approach requires only $2\times K$ operations, comprising $K$ convolutions and $K$ additions (where $K$ is the number of PSF bases). For simulated images obtained by sky survey telescopes, which often contain a large number of pixels (e.g. $9124\times 9124$), the  conventional method would require 83247376 operations. In contrast, if these images are affected by PSFs with 20 bases, our approach would necessitate only 40 operations. This significant reduction in computational complexity allows for substantial time savings when employing our principle. The whole procedure of our simulated method is shown in Figure 1, which includes two main parts: PSF bases generation part and Image Convolution and Addition part, we will discuss details of our method below.\\

\subsection{PSF Bases Generation Part} \label{p2p}

As discussed earlier, PSF bases represent a specific set that can effectively characterize PSFs across different fields of view using a limited number of coefficients. PCA is a widely recognized method for constructing PSF bases. PCA transforms high-dimensional datasets into lower-dimensional spaces while preserving the majority of the variability of the original data. This is accomplished by identifying a new set of orthogonal axes, known as principal components, along which the data points exhibit the greatest variation. In this study, we begin by obtaining a sample of PSFs to serve as our original dataset. To better capture PSF variations, we establish a sampling distance between each PSF that is smaller than the distance at which the difference between each PSF is less than $1\%$. Following this criterion, we employ the non-uniform sampling method described by \citet{jia2018ground} in our research. Typically, the number of PSF bases should exceed both the number of PCA components and the size of PSFs when represented as one-dimensional vectors. Consequently, we generally select several hundred PSFs from real observational data or generate an equivalent number using simulation code. This selection includes both the original PSFs and PSFs with random subpixel shifts to ensure a comprehensive representation of PSF variations.\\

To obtain PCA bases from the PSFs, we adopt the traditional approach outlined in~\citet{jee2007principal, nie2021constraining, sun2022precise}. First, each of these $L$ PSFs, originally an $N\times N$ pixel matrix, is transformed into a one-dimensional vector of size $P$ (where $P = N\times N$). These vectors are then stacked to form a $P\times L$ matrix. We apply Singular Value Decomposition (SVD) to this matrix, resulting in a $P\times P$ basis matrix and a $P\times L$ singular value matrix. Each column of the basis matrix represents a single PSF basis. Next, we calculate a cumulative weight figure, visualizing the cumulative proportion of variance explained by successively included principal components. The parameters for this chart are derived from the eigenvalues of the singular value matrix. Using the 'elbow' method on the cumulative weight chart, we select $K$ bases located above the elbow point \citep{jolliffe1990principal}. This method effectively differentiates signal from noise, as the contribution of meaningful components decreases gradually while random noise remains consistent. This process allows us to extract the most significant PSF components. It is worth noting that alternative methods for obtaining PSF bases can also be employed, provided they are statistically sound or physically meaningful. As long as these PSF basis are orthogonal and linearly separable, we could use the method propose below for image simulation.\\
 
\subsection{The Image Convolution and Addition Part} \label{Conv}
Having obtained the PSF bases, we can utilize them to generate simulated images from the original template image. This involves fitting the selected PSF bases to the PSFs across different fields of view, yielding coefficients that represent their contributions to the simulated images. These coefficients serve as initial values for constructing a coefficient matrix through 2D linear interpolation, accounting for the continuous variation of PSFs across the image. With $K$ PSF bases and an original image size of $N\times N$, the coefficient matrix will have dimensions $K\times N \times N$. Each element in this matrix represents the contribution of a particular PSF basis at a specific pixel location.\\ 

With PSF bases and corresponding PSF coefficient parameter matrix, we could generate simulated images from the original image by the following steps. Firstly, we will convolve the original image with all these PSF bases to generate image bases, as shown in equation~\ref{eq4}:
\begin{equation}
\label{eq4}
ImgBases_{m} = Img\ast PSFBases_{m}.
\end{equation}
In the above equation, $ImgBases_{m}$ and $PSFBases_{m}$ represent the image bases and PSF bases, respectively, for a particular index $m$. To generate the convolved image, we multiply each pixel of the $ImgBases$ with its corresponding coefficient in the coefficient matrix and sum these values for each pixel, as define below:
\begin{equation}
\label{eq5}
FImg(x,y) = \sum_{m=1}^{M} ImgBases_{m} (x,y).
\end{equation}
It is important to note that while the phase introduced by aberrations is physically continuous between pixels, the PSF itself is not. However when we consider the halo and wings of the PSF, which are not solely affected by phase variations, we need to consider these variations in spatial domain. A potential solution to this challenge is to combine PSFs derived from physical optics with halos or wings obtained from statistical models to create the final PSFs. We will further investigate this approach in our future research.\\

\section{Performance Test of The Method} \label{sec:results}
In this section, we assess the performance of our method through two distinct scenarios. First, we generate simulated data from real observation images obtained by the Ground-based Wide Angle Camera Array (GWAC), situated at the Xinglong Observatory of the National Astronomical Observatories of China (NAOC), to evaluate the fidelity of the images produced by our method \citep{xu2020real}. Second, we utilize simulated observations from the China Sky Survey Telescope (CSST) to demonstrate the enhanced computational efficiency of our approach \citep{li2024csst}. These two scenarios provide a comprehensive evaluation of both the speed improvements and image quality achieved by our proposed technique. In our paper, we utilize the traditional method to generate simulated images for comparison: first, segmenting the image into smaller patches; then, applying standard convolution functions to each segment; and finally, combining these processed patches to generate the complete blurred image. To assess performance, we measure the processing speed of this classical approach on a system featuring two Intel Xeon 6342 CPUs. Meanwhile, we evaluate our proposed method on the same computer, but with the addition of a single Nvidia RTX 3090 Ti GPU, allowing for a direct comparison of CPU-based and GPU-accelerated image blurring techniques. \\

\subsection{Simulated Image Generation Using GWAC Real Observational Data}\label{sec:lot}
Generation of simulated images using real observational data is a valuable data augmentation technique for various machine learning algorithms. The fidelity of these simulated images and the image generation speed are crucial factors. In this study, we aim to evaluate these properties comprehensively. Our real observational data were obtained from one of the cameras in the GWAC, which has a field of view of around 150 squared degree, pixel scale of 11.7 pixels/arcsec and is operating in white light mode. According to its design, PSFs in these cameras are under-sampling and may have strong spatial variations. We build a dataset spanning 170 days, comprising 15876 frames, which we assume statistically significant for our analysis. To ensure the quality and representativeness of our dataset, we carefully select images of celestial objects with a signal-to-noise ratio (SNR) larger than 10. The SNR is defined in Equation~\ref{eq6},
\begin{equation}
\label{eq6}
SNR = 10 \log_{10} \left( \frac{P_\text{signal}}{P_\text{noise}} \right),
\end{equation}
where \(P_{\text{signal}}\) represents the power of the signal and \(P_{\text{noise}}\) represents the power of the noise, respectively. Furthermore, to accurately represent PSFs across different fields of view, we make a concerted effort to include celestial objects distributed across various regions of the observational frame. However, it is important to acknowledge a potential limitation in our approach. Despite our efforts, there remains a possibility that celestial objects near the corners of the observational images may not be as well represented in our dataset. This potential under-representation could impact the comprehensiveness of our PSF modeling, particularly for PSFs that are close to the corner of the fields of view. \\

\begin{figure}[h]
\centering 
\includegraphics[width=8cm]{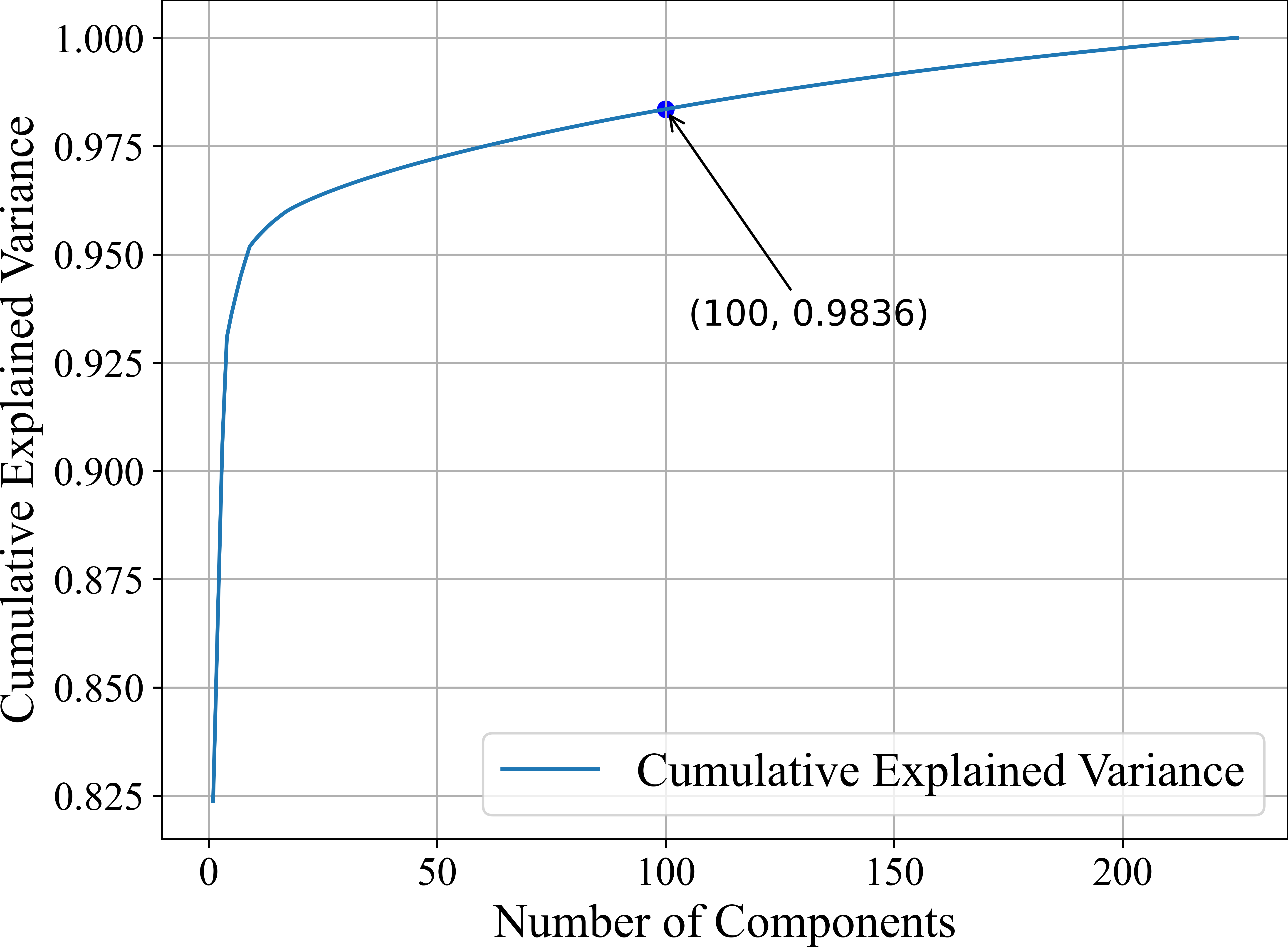} 
\caption{The method proposed in this study is delineated in this figure, which illustrates a two-stage procedure for image simulation. The first stage involves coefficient matrix generation, wherein PSF basis functions are derived from empirical PSF data. The second stage encompasses the convolution and addition step, where the previously obtained PSF basis functions are convolved with the original image to generate image components. These image components are weighted by the coefficients in each pixel and added together to produce the final blurred image. }\label{figure2}   
\end{figure}

Following the aforementioned steps, we have obtained approximately 15000 PSFs. These PSFs are initially processed using the PCA. By employing the 'elbow' method, we select 100 PCA bases, which, as illustrated in the figure~\ref{figure2}, capture over 98\% of the information contained in the original PSFs ~\ref{figure2}. These PSF bases can be utilized as basis for further simulation. To evaluate the performance of our algorithm, we choose a real observation image and extract PSF basis from it. As depicted in the figure, this image contains a diverse distribution of celestial object images. We then proceed to extract the coefficients of these PSFs with the PCA framework and apply the inverse distance weighted interpolation to obtain PCA coefficients for the entire image. Finally, we select a specific celestial object and compare the simulated image with the original image.\\

\begin{figure}[htbp]
    \centering
    \begin{minipage}[t]{0.32\textwidth}
        \centering
        \includegraphics[width=\textwidth]{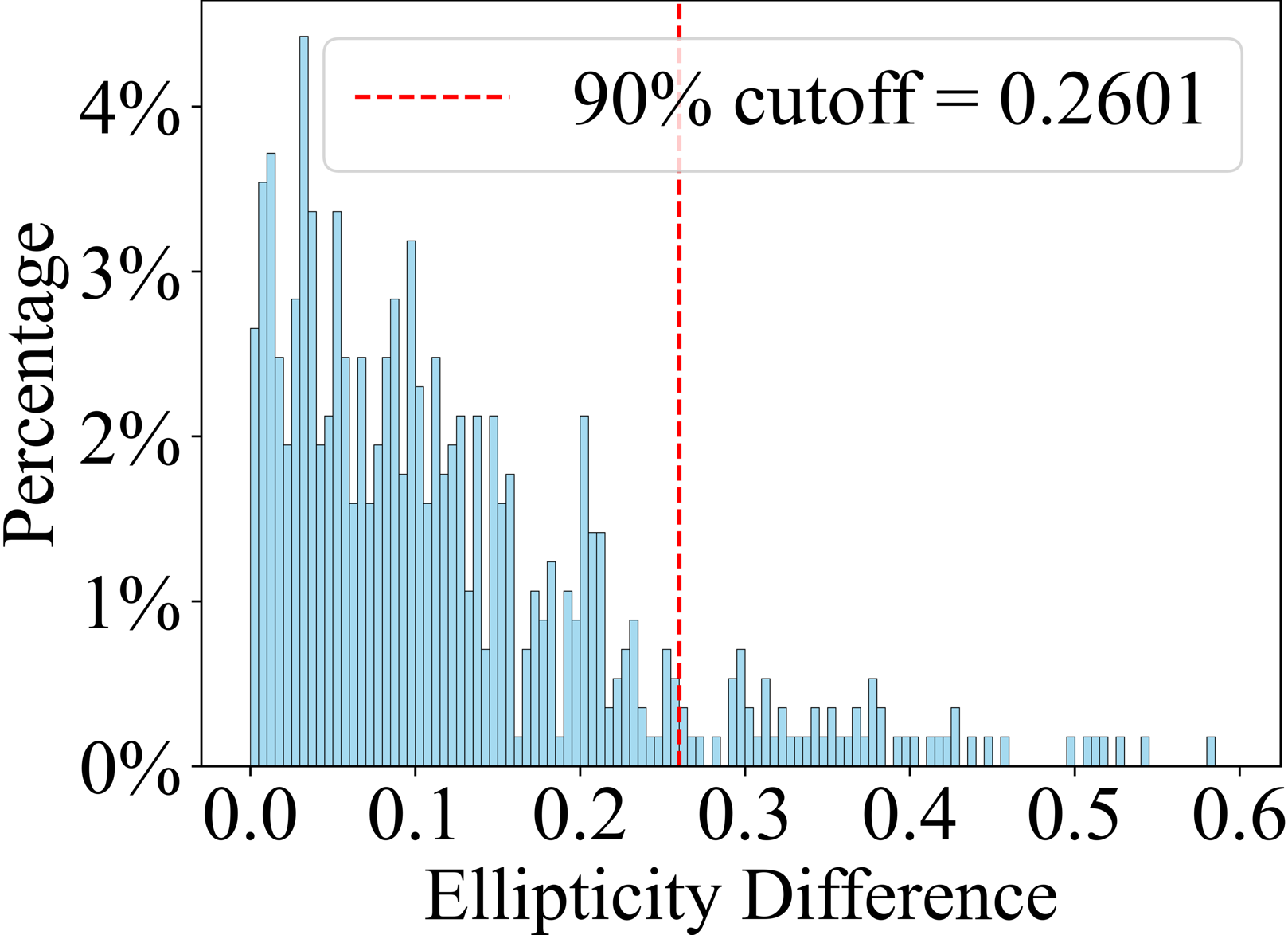}
        \textbf{(a)}
        \label{fig:gwac_ellipticity}
    \end{minipage}
    \hfill
    \begin{minipage}[t]{0.32\textwidth}
        \centering
        \includegraphics[width=\textwidth]{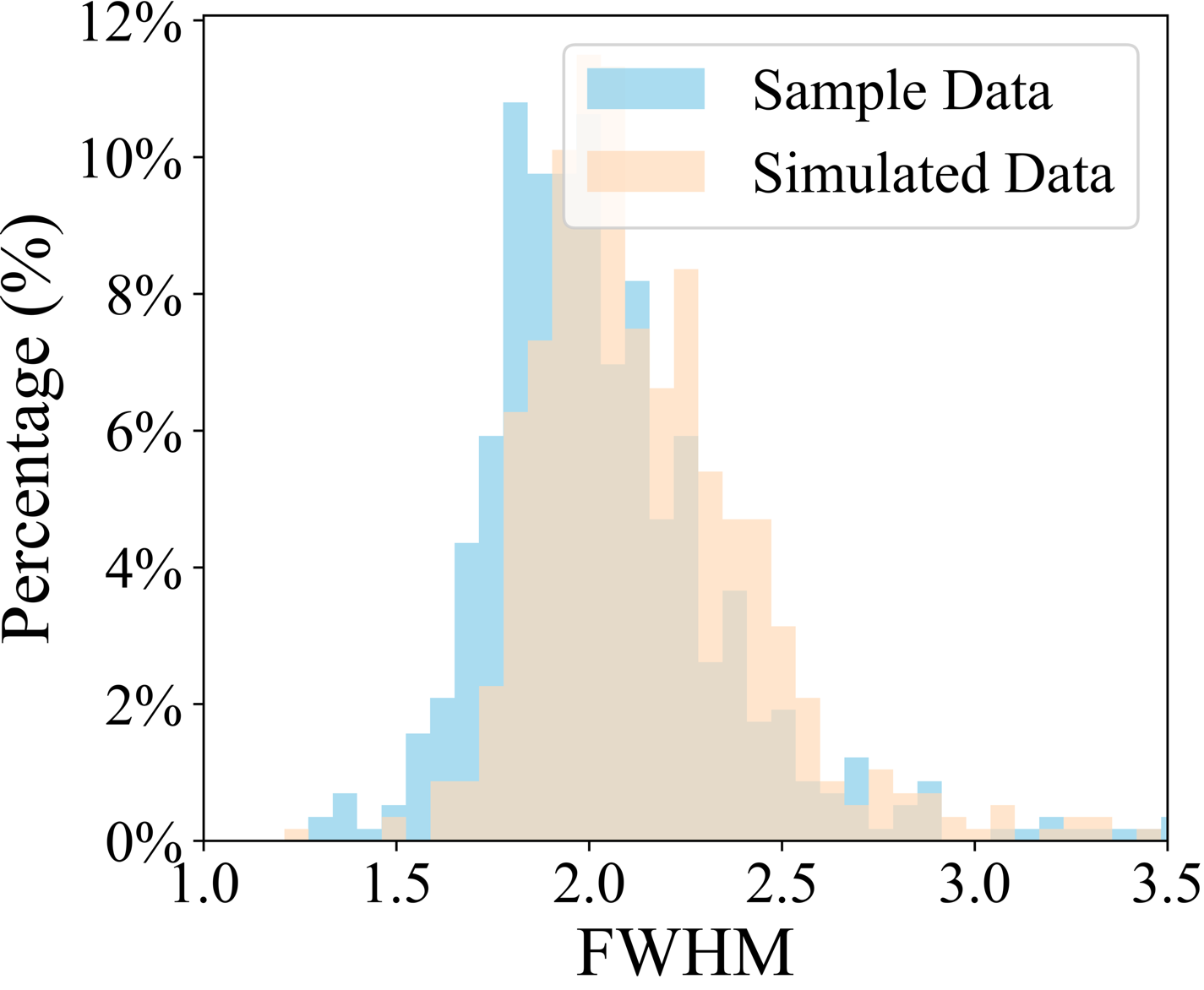}
        \textbf{(b)}
        \label{fig:FWHM_row}
    \end{minipage}
    \hfill
    \begin{minipage}[t]{0.32\textwidth}
        \centering
        \includegraphics[width=\textwidth]{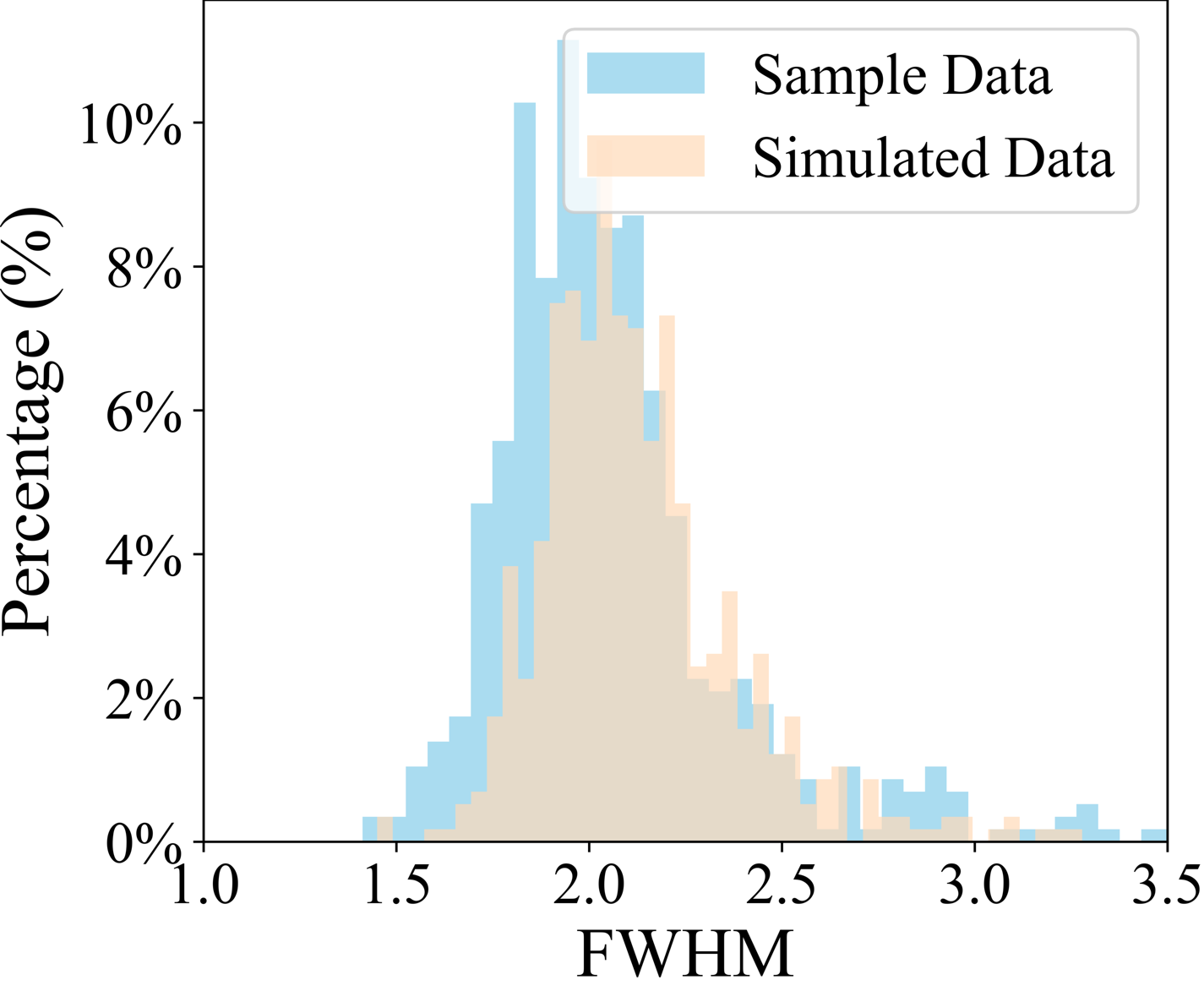}
        \textbf{( c)}
        \label{fig:FWHM_column}
    \end{minipage}
    \caption{Comparison of simulated and actual celestial object characteristics. (a) shows the histogram of the ellipticity differences between simulated images and actual celestial object images. This plot demonstrates that the simulated images have an ellipticity distribution similar to the actual images, though some differences exist due to the telescope's wide field of view and white light mode. (b) and (c) display the FWHM distribution along the row and column directions, respectively, comparing simulated data (in light orange) with sample data (in light blue). The x-axis represents the FWHM, while the y-axis represents the percentage. These histograms highlight the differences in the peak positions and spread between the simulated and actual data. These results indicate that the simulation method produces images with characteristics close to real observations.}
    \label{fig:GWAC_Ellipticity_FWHM}
\end{figure}

To obtain statistically significant results, we process 4 frames of observation data and generated 4 simulated images. It costs 1 second to generate all these simulated images. Meanwhile the traditional CPU based image simulation method will cost 12 seconds to generate 4 images. We then compare these simulated images with their original counterparts, focusing on differences in Full Width at Half Maximum (FWHM) and ellipticity. We use the following equation to calculate the FWHM:
\begin{equation}
\label{eq7}
f(x) = A \exp\left( -\frac{(x - \mu)^2}{2\sigma^2} \right),
\end{equation}
where $A$ stands for peak value of these stars and $sigma$ stands for the FWHM of these PSFs. For the ellipticity, we use the Equation~\ref{eq7} to obtain the final results~\ref{fig:GWAC_Ellipticity_FWHM}:
\begin{equation}
\label{eq8}
e = \sqrt{1 - \frac{b^2}{a^2}},
\end{equation}
where \( a \) and \( b \) represent the semi-major and semi-minor axes of the PSFs, when we fit them with elliptical equation. The results, presented in figure~\ref{fig:gwac_ellipticity}, reveal minimal discrepancies. It is worth noting that the GWAC operates in white light mode, which introduces additional complexities due to chromatic aberration and atmospheric dispersion. Considering these factors, the small differences observed between simulated and original images further validate the efficacy of our method. \\

\subsection{Simulated Image Generation According to Properties of Simulation Data}\label{sec:CSST}
To further evaluate the effectiveness of our approach, we consider a second simulation task in this section: integrating our image simulation method into a simulation framework. Specifically, we aim to incorporate our method into the simulation framework designed for the China Space Station Telescope (CSST). The CSST is a large, space-based astronomical telescope with a 2-meter aperture, offering excellent performance in terms of a large field of view and high image quality. It is equipped with the capability for on-orbit maintenance and upgrading. The CSST features five onboard instruments, including a Main Survey Camera (MSC), a Terahertz Receiver, a Multichannel Imager, an Integral Field Spectrograph, and a Cool Planet Imaging Coronagraph. The CSST comprises 30 detectors, with 18 dedicated to multi-band imaging and 12 used for spectroscopic observations. The pixel scale in the MSC is 0.074 arcseconds, and each detector captures images with $9232 \times 9216$ pixels. The MSC will use these detectors to obtain images of celestial objects.\\

Given the CSST's unprecedented capability to observe the sky with exceptional resolution and depth, generating simulated images has become challenge for testing various methods and predicting scientific outcomes. To address this need, scientists have developed specialized software to simulate images that the CSST is expected to capture \footnote{\url{https://csst-tb.bao.ac.cn/code/csst-sims/csst_msc_sim}}. The software incorporates various effects and generates PSFs for different fields of view and wavelengths, enhancing the accuracy of simulations . In the classical approach, scientists typically assume that PSFs remain uniform within very small regions. They then convolve these small regions to produce the final simulated observation images. This method generally requires 30 minutes to generate a simulated image for a single wavelength.\\

\begin{figure}
    \centering
    \includegraphics[width=8cm]{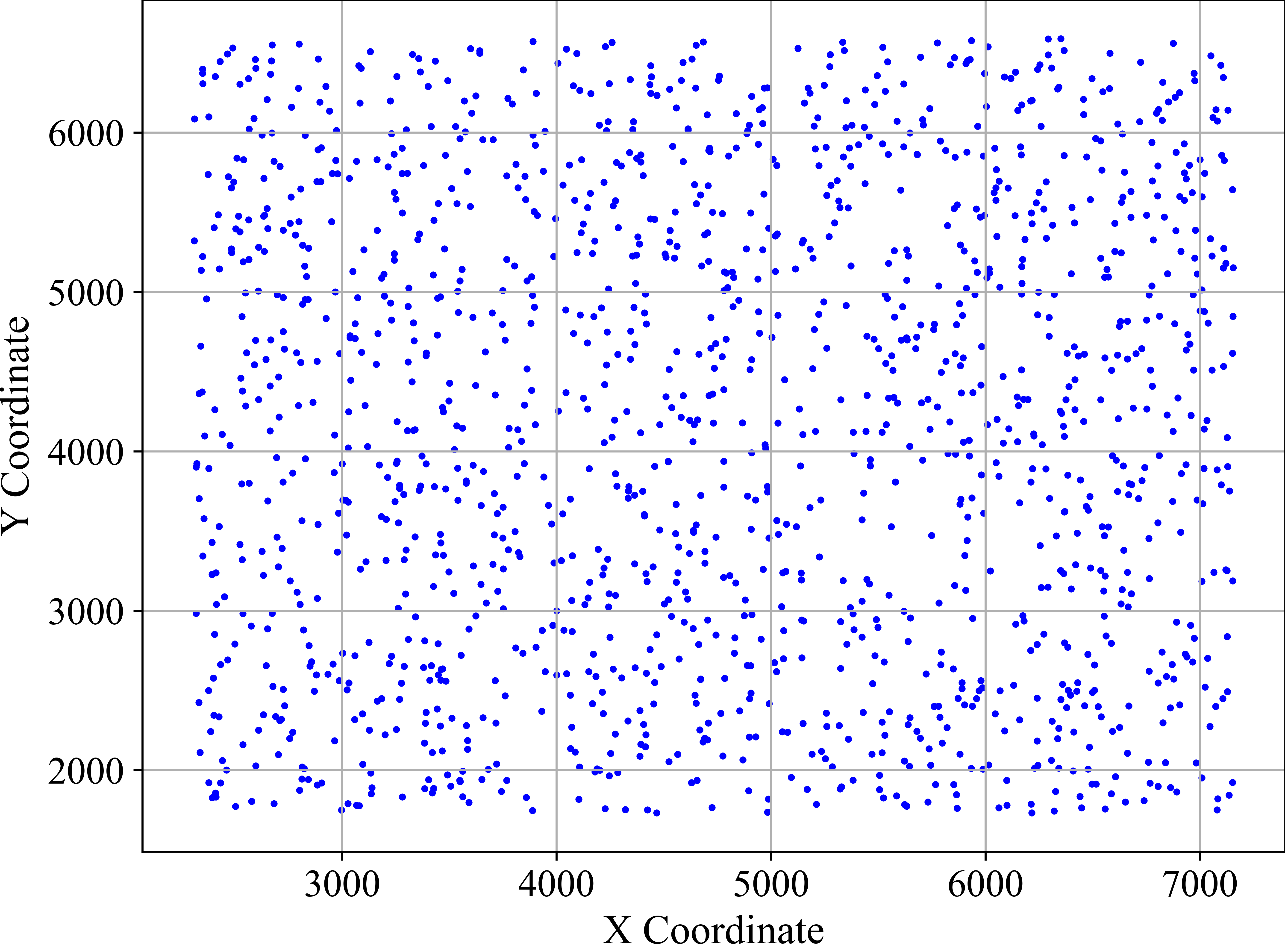}
    \caption{This figure shows the distribution of 1458 sampling points from astronomical images, where duplicate coordinates have been removed and the remaining points were randomly selected. The X-axis and Y-axis represent the image’s horizontal and vertical coordinates, respectively. The blue dots represent the positions of the sampled points.}
    \label{figure4}
\end{figure}

In our approach, we begin by sampling PSFs across one of the detectors using the previously described method. As illustrated in the figure~\ref{figure4}, we have obtained 5041 PSFs distributed throughout the entire detector. Subsequently, we decompose these PSFs into 100 PSF bases and employ our novel method to generate simulated images. The process requires approximately 9.51 seconds on average to produce a single frame of simulated images. Notably, the image size and the number of celestial objects directly influence the computational load. To account for this, we have generated simulated images with varying dimensions and different quantities of celestial objects, as detailed in the accompanying table~\ref{table1}.\\

\begin{table}[h]
\centering
\renewcommand{\arraystretch}{1.2}
\begin{tabular}{|c|ccccc|ccccc|}
\hline
\multirow{1}{*}{Original Image Size}                & \multicolumn{5}{c|}{\multirow{1}{*}{$512\times 512$ pixels}}                                                                                          & \multicolumn{5}{c|}{\multirow{1}{*}{$1024\times 1024$ pixels}}                                                                                          \\ \hline
\multirow{2}{*}{Different Methods}   & \multicolumn{4}{c|}{\multirow{1}{*}{\makebox[0.24\textwidth][c]{Classical Method (Different Patch Size)}}}                                                         & \multirow{2}{*}{Ours} & \multicolumn{4}{c|}{\multirow{1}{*}{\makebox[0.24\textwidth][c]{Classical Method (Different Patch Size)}}}                                                           & \multirow{2}{*}{Ours} \\ \cline{2-5} \cline{7-10}
                          & \multicolumn{1}{c|}{\multirow{1}{*}{\makebox[0.06\textwidth][c]{$32\times 32$}}}     & \multicolumn{1}{c|}{\multirow{1}{*}{\makebox[0.06\textwidth][c]{$64 \times 64$}}}    & \multicolumn{1}{c|}{\multirow{1}{*}{\makebox[0.06\textwidth][c]{$128\times 128$}}}   & \multicolumn{1}{c|}{\multirow{1}{*}{\makebox[0.06\textwidth][c]{$256\times256$}}}  &                       & \multicolumn{1}{c|}{\multirow{1}{*}{\makebox[0.06\textwidth][c]{$32\times 32$}}}     & \multicolumn{1}{c|}{\multirow{1}{*}{\makebox[0.06\textwidth][c]{$64 \times 64$}}}    & \multicolumn{1}{c|}{\multirow{1}{*}{\makebox[0.06\textwidth][c]{$128\times 128$}}}   & \multicolumn{1}{c|}{\multirow{1}{*}{\makebox[0.06\textwidth][c]{$256\times256$}}}   &                       \\ \hline
\multirow{1}{*}{Average Time Cost (s)} & \multicolumn{1}{c|}{\multirow{1}{*}{206.94}} & \multicolumn{1}{c|}{\multirow{1}{*}{55.05}} & \multicolumn{1}{c|}{\multirow{1}{*}{17.46}} & \multicolumn{1}{c|}{\multirow{1}{*}{7.39}} & {\multirow{1}{*}{1.24}}                  & \multicolumn{1}{c|}{\multirow{1}{*}{807.73}} & \multicolumn{1}{c|}{\multirow{1}{*}{204.55}} & \multicolumn{1}{c|}{\multirow{1}{*}{57.47}} & \multicolumn{1}{c|}{\multirow{1}{*}{19.34}} & {\multirow{1}{*}{2.87}}                  \\ \hline
\end{tabular}
\caption{The total time cost of different methods in generation of blurred images. As shown in this table, our method is much faster than the classical method, even when we use patches with relatively large size in the classical method.}
\label{table1}
\end{table}

\begin{figure}[htbp]
    \centering
    \begin{minipage}[t]{0.46\textwidth}
        \centering
        \includegraphics[width=\textwidth]{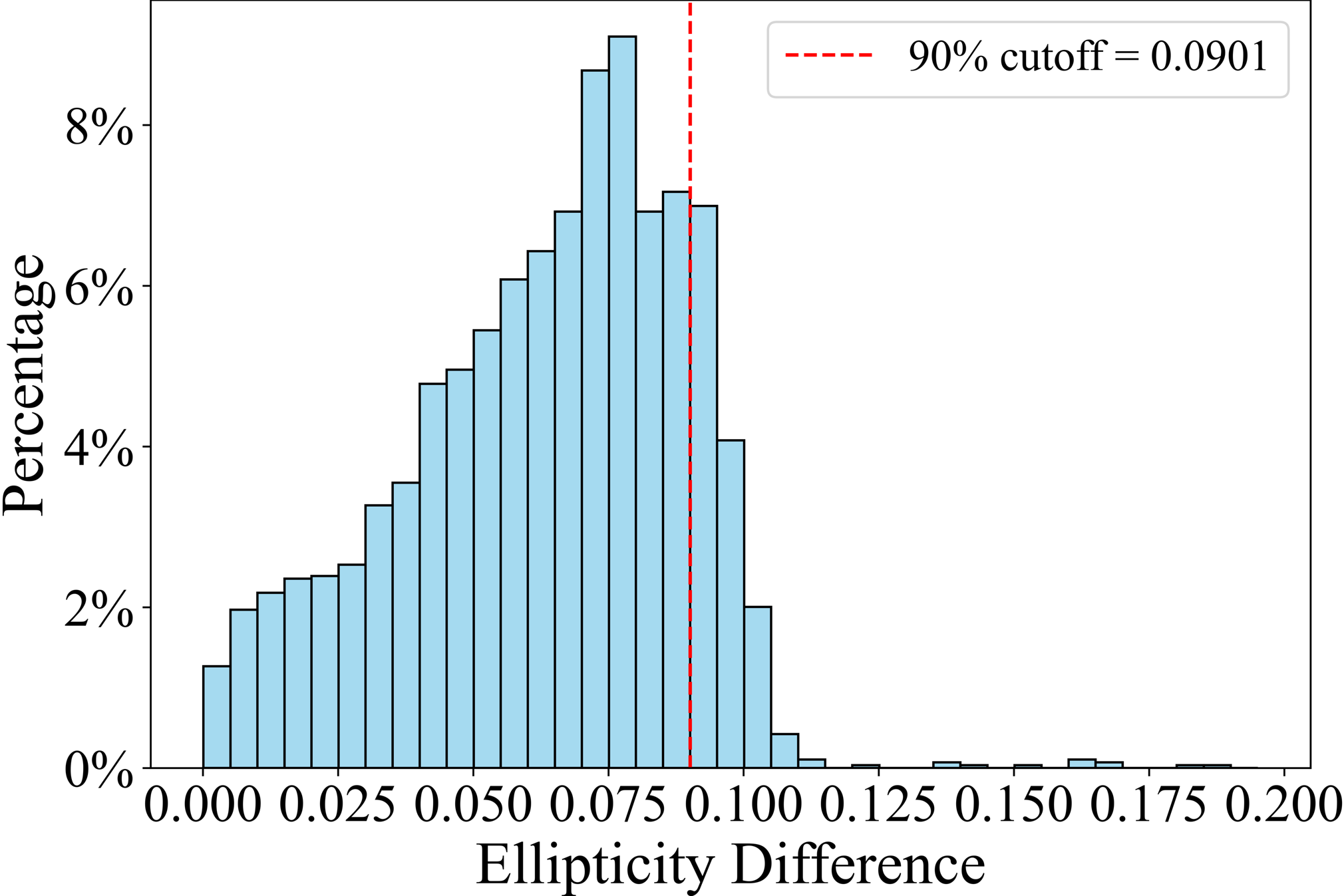}
        \textbf{(a)}
    \end{minipage}
    \hfill
    \begin{minipage}[t]{0.46\textwidth}
        \centering
        \includegraphics[width=\textwidth]{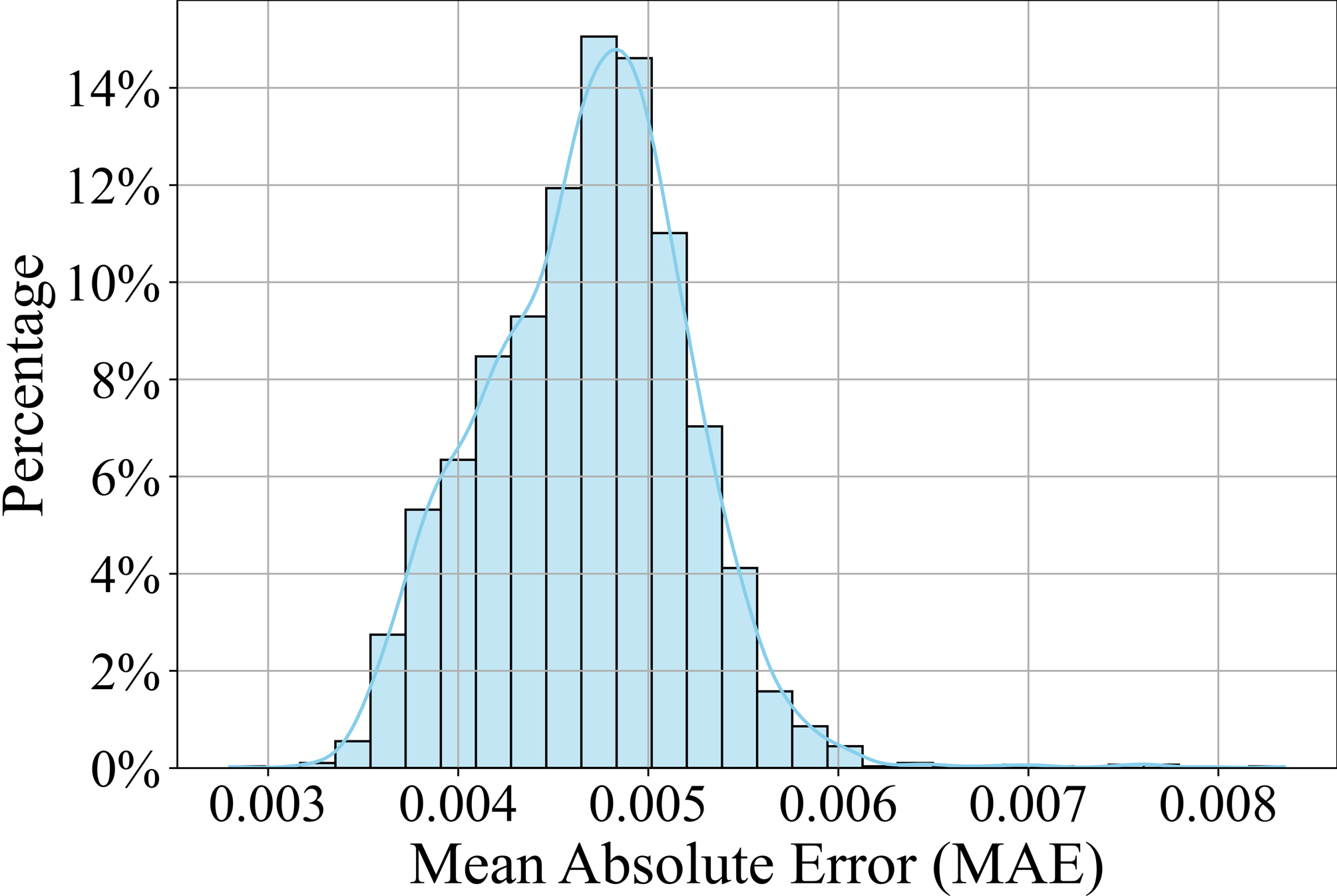}
        \textbf{(b)}
    \end{minipage}
    \caption{Comparison of two key metrics between simulated and true PSFs. (a) displays the distribution of ellipticity differences, with the x-axis representing the ellipticity difference and the y-axis indicating the percentage of occurrence for each value. A red dashed line marks the 90\% cutoff point at approximately 0.0901, identifying the range where most ellipticity differences are concentrated. (b) illustrates the distribution of Mean Absolute Error (MAE) values across 2916 positions, with the x-axis showing the MAE values and the y-axis representing the percentage of occurrence for each value. Together, these plots provide insights into the variation and concentration of differences between simulated and true PSFs.}
    \label{fig:CSST_Ellipticity_MAE}
\end{figure}

\begin{figure*}[htbp]
    \centering
    \begin{minipage}[t]{0.3\textwidth}
        \centering
        \includegraphics[width=\textwidth]{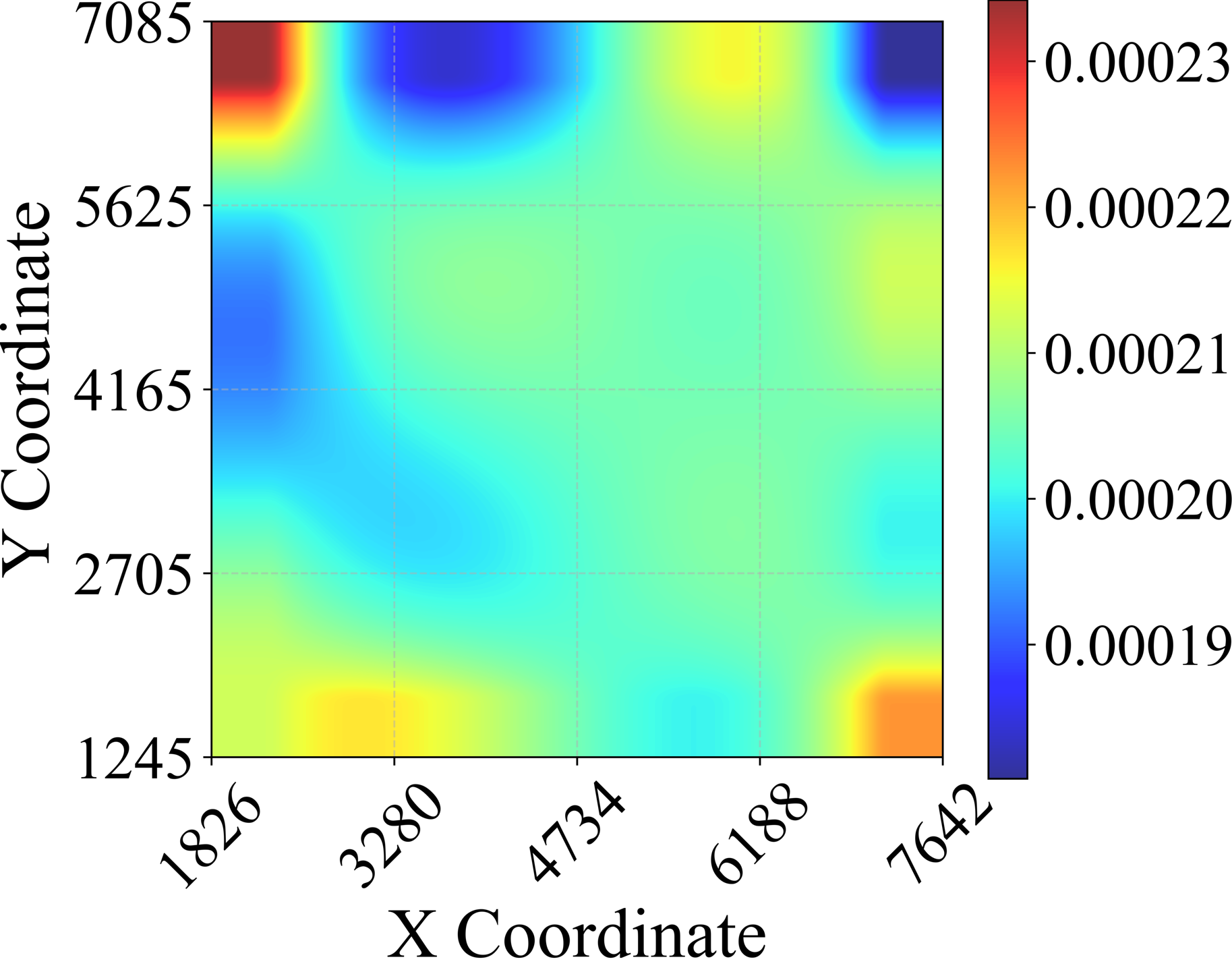}
        \textbf{(a)}
        \label{fig:MSE_Scatter}
    \end{minipage}
    \hfill
    \begin{minipage}[t]{0.6\textwidth}
        \centering
        \includegraphics[width=0.49\textwidth]{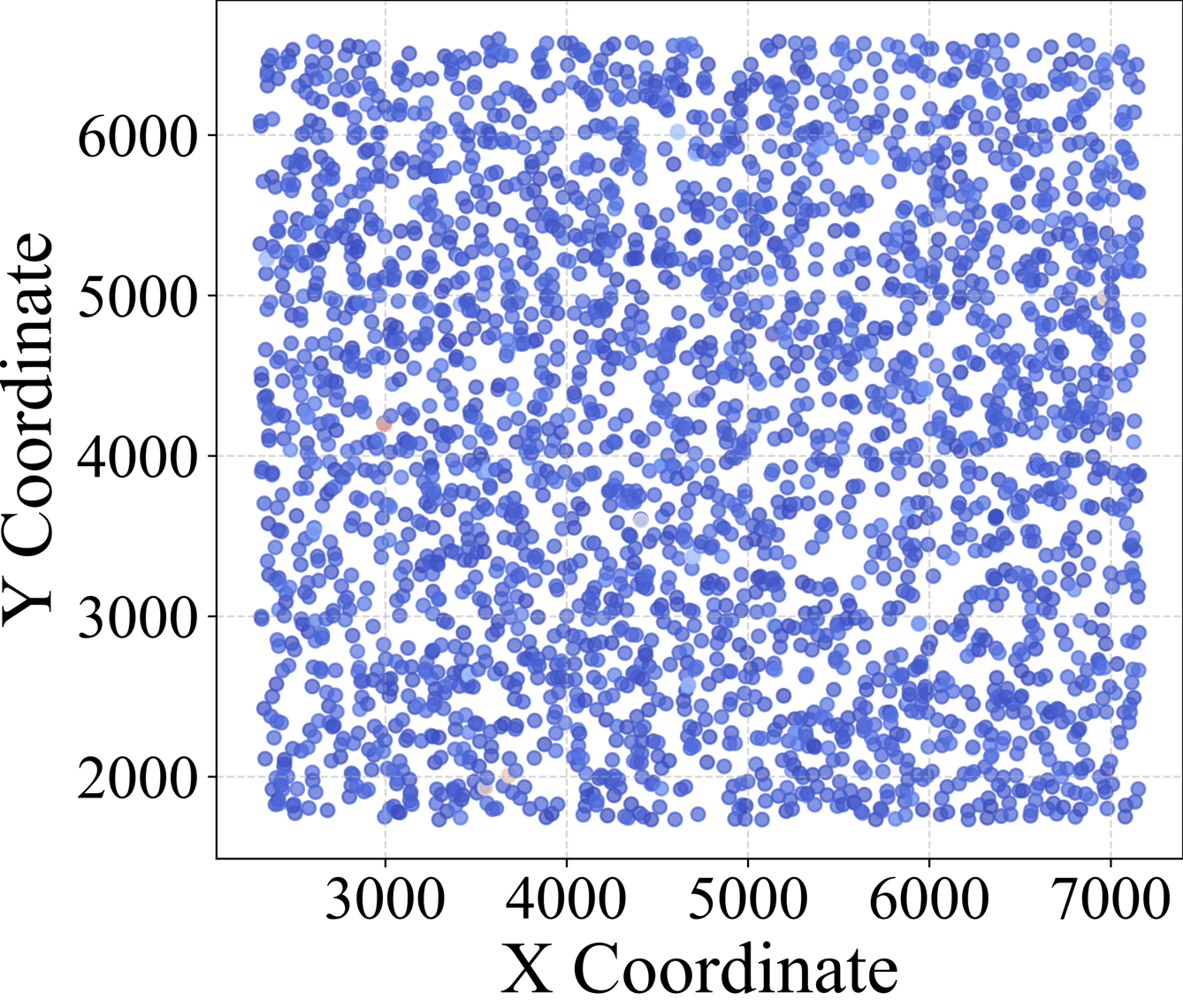}
        \includegraphics[width=0.49\textwidth]{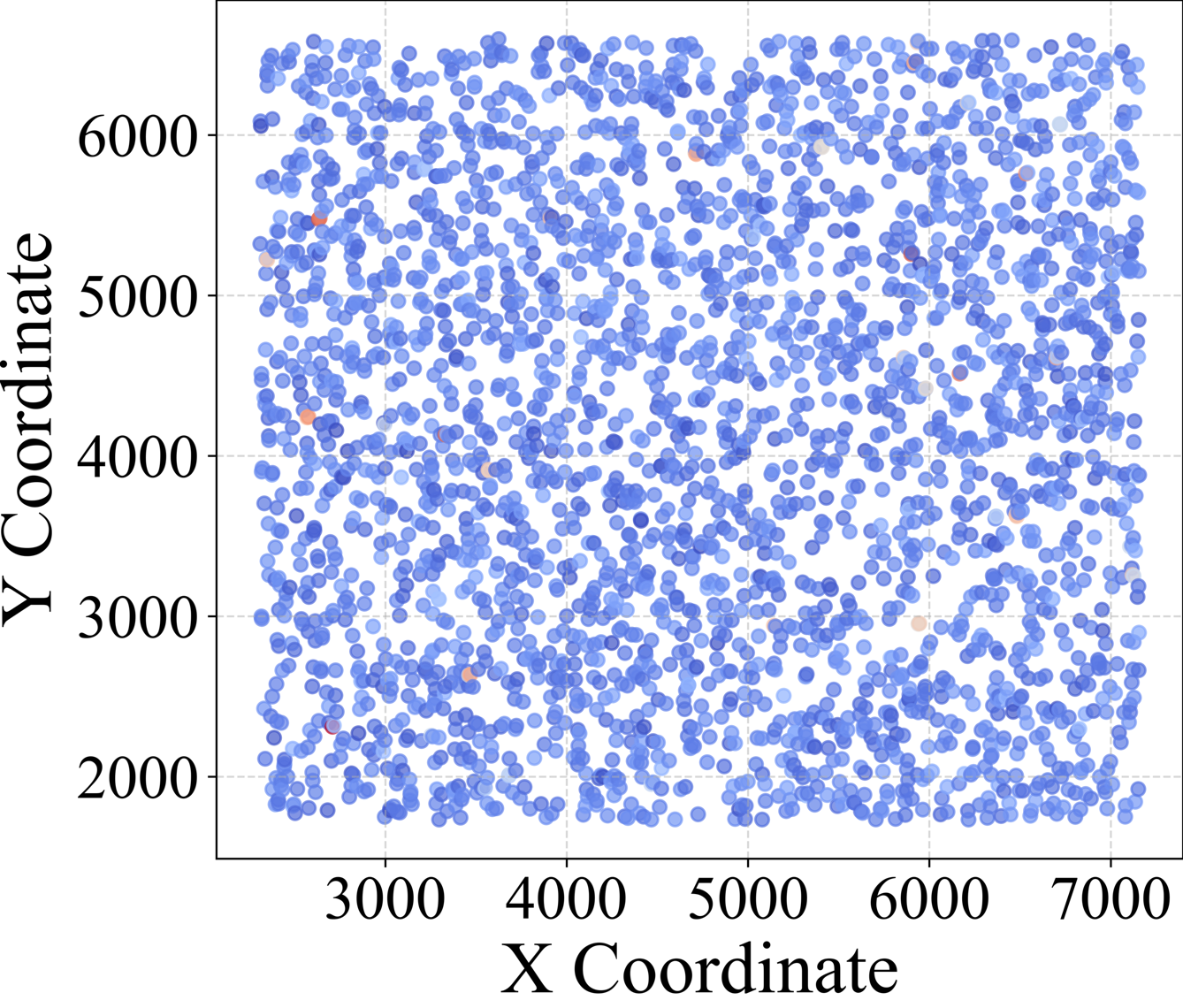}

        \textbf{(b)}
        \label{fig:ellipticity_difference_scatter}
    \end{minipage}
    \begin{minipage}[t]{0.04\textwidth}
        \centering
        \raisebox{0.4cm}{ 
            \includegraphics[width=\textwidth]{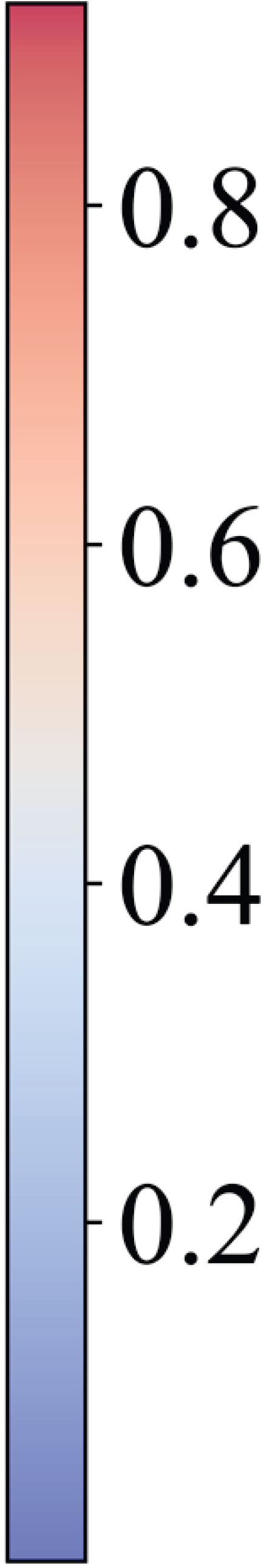}
        }
    \end{minipage}
    \caption{Spatial comparison of Mean Squared Error (MSE) and ellipticity component differences between simulated and true data. (a) presents the MSE distribution across a 2D grid, highlighting error variations across different areas. (b) shows the spatial differences in ellipticity components $e_1$ (left) and $e_2$ (right), with both subplots using a unified color scale to represent the magnitude of these differences. The left subplot displays the distribution of differences in the $e_1$ component, while the right subplot displays the distribution of differences in the $e_2$ component. Together, these visualizations provide insights into the spatial accuracy and ellipticity consistency of the simulation compared to actual data.}
    \label{fig:MSE_and_ellipticity_difference}
\end{figure*}

\begin{figure}[h]
\centering 
\includegraphics[width=13.5cm]{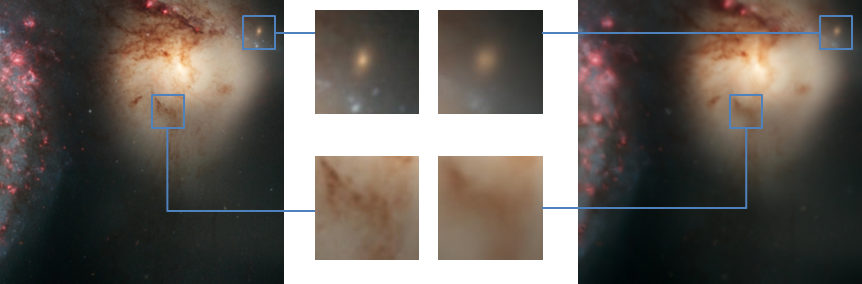} 
\caption{The original and blurred image obtained by our method. We can find that there are no boundaries in the blurred image, which could often be seen in images generated by other methods.}\label{figure_simulate} 
\end{figure}

To evaluate the accuracy of our simulated images, we have generated a series of images, each containing 2916 celestial objects, using both the classical method and our proposed approach. We have then calculated the Mean Absolute Error - MAE (\(  \frac{1}{n} \sum_{i=1}^{n} | y_i - \hat{y}_i | \)) and ellipticity of these PSFs. The results of this analysis are presented in Figure~\ref{fig:CSST_Ellipticity_MAE}. As evident from this figure, our method demonstrates the capability to generate highly realistic images with minimal discrepancies. Additionally, we have created a heatmap to visualize the distribution of residual errors. Figure~\ref{fig:MSE_and_ellipticity_difference} (a) and figure~\ref{fig:MSE_and_ellipticity_difference} (b) illustrate that our method successfully generates simulated images with the majority of differences primarily concentrated around 0. Finally, we present a simulated image of a nearby galaxy to showcase the capabilities of our method in figure~\ref{figure_simulate}. Given the substantial size of nearby galaxies, traditional simulation methods often produce noticeable boundaries between image segments. In contrast, our approach generates realistic simulated images without these artificial boundaries, as demonstrated in the figure~\ref{figure_simulate}. This seamless integration highlights the superior performance of our method in creating large-scale astronomical simulations, particularly for extended celestial objects.\\


\section{Conclusions and Future Works} \label{sec:con}
The advancement of diverse astronomical instruments and technologies necessitates high-fidelity simulated images with spatially and temporally variable PSFs. Classical methods, however, are computationally demanding and often produce blurred images with discernible boundaries. This paper introduces an innovative approach for generating high-fidelity simulated images. Our method enables the creation of images with spatially variable PSFs and seamless boundaries. This technique has potential applications in the development and performance evaluation of various telescopes. Moreover, it facilitates the generation of large-scale simulated image datasets, which are particularly crucial for training diverse artificial intelligence algorithms in astronomical research and data processing.\\

Our analysis reveals residual discrepancies between images generated by our method and those produced through classical approaches. These differences may stem from the PCA decomposition process, suggesting the need for further investigation into novel techniques for deriving more effective PSF bases tailored to our specific applications. Additionally, given that both our method and the classical approach are grounded in PSF concepts, it is crucial to incorporate high-fidelity PSF simulation or modeling techniques. These enhancements should account for not only diffraction effects but also rings and halos, thereby producing more realistic results. Moreover, we plan to explore methods for integrating chromatic aberration and atmospheric dispersion effects, further improving the authenticity of our simulated images.\\

\section*{acknowledgement}
The code used in this paper will be shared in the PaperData Repository powered by the China-VO. This work is supported by the National Key R \& D Program of China (No. 2023YFF0725300) and the National Natural Science Foundation of China (NSFC) with funding numbers 12173027. This work is supported by the Young Data Scientist Project of the National Astronomical Data Center. \\

\section*{Data Availability}
The code and data used in this paper would be released in PaperData Repository powered by China-VO with DOI number of 10.12149/101525.\\


\bibliography{sample631}{}
\bibliographystyle{aasjournal}



\end{document}